\documentclass[aps,showpacs,twocolumn,prd,superscriptaddress,nofootinbib]{revtex4}

\usepackage{amsmath}
\usepackage{latexsym}
\usepackage{graphicx}
\usepackage{color}
\usepackage{enumerate}
\usepackage{ulem}

\newcommand{\beq}{\begin{equation}}
\newcommand{\eeq}{\end{equation}}
\newcommand{\bea}{\begin{eqnarray}}
\newcommand{\eea}{\end{eqnarray}}
\newcommand{\ba}{\begin{array}}
\newcommand{\ea}{\end{array}}
\newcommand{\MSun}{{\rm M}_{\odot}}

\newcommand{\Abar}{\bar{A}}
\newcommand{\Ebar}{\bar{E}}
\newcommand{\Tbar}{\bar{T}}

\newcommand{\hB}{h_B}

\newcommand{\eg}{e.~g.~}
\def\leq{\,\raise 0.4ex\hbox{$<$}\kern -0.8em\lower 0.62ex\hbox{$-$}\,}
\def\geq{\,\raise 0.4ex\hbox{$>$}\kern -0.8em\lower 0.62ex\hbox{$-$}\,}
\def\pm{\,\raise 0.4ex\hbox{$+$}\kern -0.8em\lower 0.62ex\hbox{$-$}\,}

\begin{document}

\title{Sky localization of complete inspiral-merger-ringdown signals for nonspinning massive black hole binaries}

\author{Sean T. McWilliams}
\email{sean@astro.columbia.edu}
\affiliation{Gravitational Astrophysics Laboratory, NASA Goddard Space Flight Center, 8800 Greenbelt Rd., Greenbelt, MD 20771, USA}
\affiliation{Institute for Strings, Cosmology and Astroparticle Physics (ISCAP), Columbia University, New York, NY 10027, USA}
\affiliation{Physics Department, Princeton University, Princeton, NJ 08544, USA}
\author{Ryan N. Lang}
\affiliation{Gravitational Astrophysics Laboratory, NASA Goddard Space Flight Center, 8800 Greenbelt Rd., Greenbelt, MD 20771, USA}
\author{John G. Baker}
\affiliation{Gravitational Astrophysics Laboratory, NASA Goddard Space Flight Center, 8800 Greenbelt Rd., Greenbelt, MD 20771, USA}
\author{James Ira Thorpe}
\affiliation{Gravitational Astrophysics Laboratory, NASA Goddard Space Flight Center, 8800 Greenbelt Rd., Greenbelt, MD 20771, USA}
\date{\today}

\begin{abstract}

We investigate the capability of LISA to measure the sky position of equal-mass, nonspinning black hole binaries, combining for the first time the entire inspiral-merger-ringdown signal, 
the effect of the LISA orbits, and the complete three-channel LISA response.  We consider an ensemble of systems near the peak of LISA's sensitivity band, with total rest mass of $2 \times 10^6\ \MSun$, a redshift of $z=1$, and randomly chosen orientations and sky positions.  We find median sky localization errors of approximately $\sim3$ arcminutes. This is comparable to the field of view of powerful 
electromagnetic telescopes, such as the James Webb Space Telescope, that could be used to search for electromagnetic signals associated with merging massive black holes. We investigate the way in which parameter 
errors decrease with measurement time, focusing specifically on the additional information provided during the merger-ringdown segment of the signal. We find that this information improves all parameter 
estimates directly, rather than through diminishing correlations with any subset of well-determined parameters.  Although we have employed the baseline LISA design for this study, many of our conclusions
regarding the information provided by mergers will be applicable to alternative mission designs as well.
\end{abstract}


\maketitle

\section{Introduction}
\label{sec:intro}

Among the gravitational-wave (GW) sources which the proposed Laser Interferometer Space Antenna (LISA) would observe,
massive black hole binaries (MBHBs) stand out in several
ways.  They have the highest typical signal-to-noise ratios (SNRs) and will be observable out to redshifts of $z\sim10$ or greater
\cite{Flanagan:1997sx,Baker:2006kr,McWilliams:2010},
making MBHBs the most distant sources that LISA will observe.
Observations of MBHBs figure prominently
in LISA's science goals \cite{LISASciCase} and may provide particularly 
rich opportunities for multi-messenger observation in conjunction with
electromagnetic instruments.  The potential for such observations
depends particularly on LISA's ability to localize the gravitational-wave
sources on the sky and thereby facilitate simultaneous or follow-up electromagnetic observation \cite{Kocsis:2007yu}.

LISA is an all-sky instrument in the sense that, at any one instant, it is sensitive to sources at most points on the sky. While this is advantageous from the perspective of detection of unknown sources, it is not helpful for locating these sources on the sky. To build localization information, LISA takes advantage of the fact that MBHBs are observable for months or even years prior to merger. During this time, LISA's orbit around the Sun introduces both frequency and amplitude modulations to the signal that can be used to determine source location \cite{Cutler:1997ta}. The majority of studies of LISA's source location abilities \cite{Cutler:1997ta, Lang:2006bz, Kocsis:2007hq, Arun:2007hu, Trias:2008prd, Lang:2007ge, Porter:2008kn, Lang:2008gh, Klein:2009, Lang:2011je} focus on this mechanism and neglect the final merger-ringdown portion of the MBHB waveform, which is difficult to model and too brief to experience any modulation from LISA's orbit.

Recent work \cite{Thorpe:2008wh, McWilliams:2009pe} has shown that the
merger-ringdown portion of the MBHB signal, which is short-lived but
has more power and occurs at higher-frequencies, can achieve sky
localization comparable to the long inspiral portion without any
additional information from LISA's orbital modulation.
The mechanism for extracting this information is believed to be related to the complex response of LISA to gravitational waves at the upper end of the measurement band (above $\sim3$ mHz, where finite arm length effects begin to become important \cite{Rubbo:2004}).  In this paper, we investigate the combined effect of both localization mechanisms.  Specifically, we expand our previous analysis to include the full inspiral signal, including orbital effects.  We also include a third signal channel, $\Tbar$, relevant only at higher frequencies, which was excluded from the earlier analysis for technical reasons.

The outline of the paper is as follows: In Section~\ref{sec:meth}, we review
our methodology, which closely follows \cite{McWilliams:2009pe}.  Section~\ref{subsec:waveform} describes the construction of the inspiral-merger-ringdown (IMR) waveform using the ``IRS-EOB'' model \cite{Baker:2008mj}.  Section~\ref{subsec:instrument} describes our model for the LISA instrument, including the response function and noise.  Finally, Section~\ref{subsec:errorEst} explains how we calculate parameter errors using the standard Fisher matrix formalism.

In Section \ref{sec:res}, we examine LISA's ability to localize
sources, including the improvements that are gained as different levels of realism are added.  We find that for an ensemble of systems with total rest mass of $2 \times 10^6\ \MSun$ at a redshift of $z=1$, observing the 
IMR signal with a stationary detector (and
therefore without the benefit of orbital modulation) provides better sky localization than observing the inspiral signal with an orbiting detector. 
We then consider the observation of a complete IMR signal with an orbiting detector and find the final error to be only a
modest improvement over the IMR signal-stationary detector result.  Measuring the long-duration inspiral is essentially irrelevant to achieving LISA's localization potential.  Nevertheless, the orbital information contained in the inspiral is still important for localizing signals in advance of merger, which may be critical for conducting counterpart searches. We find, in agreement with previous results, that LISA can localize most MBHBs to within the Large Synoptic Survey Telescope (LSST) field of view (FOV) several months before merger.  With the complete IMR signal, $\sim50\%$ of MBHBs in our ensemble can be localized into the much smaller FOV of the James Webb Space Telescope (JWST).  This is possible mainly because of the extra information provided by the $\Tbar$ channel.

In Section~\ref{sec:discussion}, we demonstrate that the merger-ringdown (MR) portion of the waveform provides additional information beyond the 
inspiral for all parameters. 
We summarize and conclude in Section~\ref{sec:conc}, focusing on how our results may impact alternative designs for LISA.

\section{Methodology}
\label{sec:meth}

The procedure used to estimate the errors with which LISA will measure astrophysical source parameters of MBHBs closely follows that of our previous studies \cite{Thorpe:2008wh, McWilliams:2009pe}. We will provide a brief overview; those interested in more detail should consult \cite{McWilliams:2009pe}.  To assess parameter accuracy, we require a model for the emitted waveform, a model for the detector response and noise,
and a method for converting these pieces of information into a theoretical limit on the achievable accuracy in measuring parameters with LISA.

\subsection{Waveform Model}
\label{subsec:waveform}

We use a complete IMR waveform model \cite{Baker:2008mj} tuned to match the available
numerical simulations for nonspinning black hole binaries.  This model, referred to as the IRS-EOB model, 
uses a conventional effective-one-body (EOB) Hamiltonian formalism to model the adiabatic inspiral \cite{Buonanno:2007pf}.  The merger-ringdown is modeled using the ``implicit rotating source'' (IRS) formalism, which treats the radiation source as a shrinking rigid rotator \cite{Baker:2008mj}. 

The waveform amplitude is calculated using a flux model that is constrained both to be consistent with the inspiral flux through 3.5 post-Newtonian (PN) order and also to vanish as it approaches the ringdown frequency (referred
to as ``Model 2'' and given by Eq.~19 in \cite{Baker:2008mj}). The waveform phase is fit to a physically motivated functional form (see Eq.~9 
in \cite{Baker:2008mj}).

Working in geometrized units where $G=c=M=m_1+m_2=1$, the source model depends only on the mass ratio $q\equiv m_1/m_2$ (where $m_1<m_2$) and the
spins.
We generate a waveform time series using the IRS-EOB model with a signal cadence
corresponding to a quarter wavelength at the highest frequency reached by the $\ell=4$, $m=\pm4$ harmonics\footnote{We oversample to accommodate interpolation
when applying the response.}.  After the source calculation, we transform the waveform from the source coordinates to the solar system barycenter (SSB).  The SSB waveform $\hB$ is given by:
\beq
\hB = \frac{GM}{c^2D_L}\left[e^{2i\psi}\sum_{\ell m}\, ^{-2}Y_{\ell m}(\iota,\phi_o)\,h_{\ell m}\left(\frac{t_c-t}{M}\right)\right]\,,
\label{eq:source2SSB}
\eeq
where $^{-2}Y_{\ell m}$ are the spherical harmonics of spin-weight $-2$ \cite{Goldberg:1967} and $h_{\ell m}$ is the dimensionless strain
decomposed in that basis in the source frame.  This transformation depends on six
parameters: the redshifted total system mass $M=M_o(1+z)$, the 
luminosity distance $D_{L}$, the coalescence time $t_{c}$, and three angles describing
the orientation of the binary.  The coalescence time, $t_c$, is defined as the time at which the peak of the waveform reaches the solar system barycenter.  For the angles, we use the inclination $\iota$ 
(with the convention that $\iota=0$ corresponds to the line of sight 
being coincident with the orbital axis of the binary),
initial orbital phase $\phi_{o}$, and polarization phase $\psi$.  

We focus on a specific case that has appeared elsewhere in the literature (\eg \cite{Kocsis:2007hq,Lang:2006bz}):
an equal-mass, nonspinning MBHB with rest mass $M_o=2\times 10^6\ \MSun$ at $z=1$, observed for the final year prior to merger.  This mass is interesting because it lies near the most sensitive range of the LISA instrument \cite{McWilliams:2010}, merger-tree models predict LISA will observe a significant number of events in this range \cite{Sesana:2007sh}, and excellent observational evidence exists for black holes of this size \cite{Gillessen:2009,Ghez:2005,Barth:2004fw}.

The sky location of the binary, described by the ecliptic latitude $\beta$ and longitude $\lambda$,
is applied by the instrument response, which we discuss below.
We use the vector
$\Lambda^a\equiv(\ln M,\ln D_{L},\beta,\lambda,\iota,\phi_{o},\psi,t_{c})$ to
denote the complete set of variable parameters.  We exclude the mass ratio $q$ from $\Lambda^a$, as it is not 
varied when computing parameter uncertainties, a procedure consistent with that used in \cite{Babak:2008bu,Thorpe:2008wh,McWilliams:2009pe}. 

\subsection{Instrument Model}
\label{subsec:instrument}

The instrument model includes the transformation of the SSB waveform $\hB$ to the LISA instrument outputs as well as a model of the instrument noise. The instrument outputs are the \{$\Abar$, $\Ebar$, $\Tbar$\} TDI combinations described in \cite{McWilliams:2009pe} and based on the
orthogonal \{$A$, $E$, $T$\} combination \cite{Prince:2002hp}. The response is computed by employing {\sc Synthetic LISA} \cite{Vallisneri:2004bn}
to generate the \{$X$, $Y$, $Z$\} TDI combinations \cite{Tinto:2003uk} and then converting to \{$\Abar$, $\Ebar$, $\Tbar$\} using
\bea
\Abar&=& \frac{Z-X}{2\sqrt{2}} \,, \nonumber \\
\Ebar&=& \frac{X+Z-2\,Y}{2\sqrt{6}} \,, \nonumber \\
\Tbar&=& \frac{X+Y+Z}{2\sqrt{3}}\,.
\eea

We use an analytic model of the noise in the \{$\Abar$, $\Ebar$, $\Tbar$\} TDI channels, obtained following the procedure in \cite{Estabrook:2000ef}. The expressions for the one-sided spectral densities are
\begin{widetext}
\bea
S_{\bar A, \bar E}&=&2\sin^2(\Phi)\left[2\left(3+2\cos(\Phi)+\cos(2\,\Phi)\right)S_{\rm pm}+\left(2+\cos(\Phi)\right)S_{\rm op}\right], \nonumber \\
S_{\Tbar}&=&8\sin^2(\Phi)\sin^2(\Phi/2)\left[4\sin^2(\Phi/2)S_{\rm pm} + S_{\rm op}\right]\,,
\label{eq:Saet}
\eea
\end{widetext}
where $\Phi \equiv 2\pi f L/c$, $f$ is the frequency, 
$L$ is the arm length, 
and $S_{\rm pm}$ and $S_{\rm op}$ are the one-sided
spectral densities of the proof-mass acceleration and optical path-length noises, 
\bea
S_{\rm pm} &=& 2.5\times 10^{-48}\left( \frac{f}{1\,\mbox{Hz}}\right)^{-2}\sqrt{1+\left(\frac{f}{0.1\,\mbox{mHz}}\right)^{-2}}, \nonumber \\
S_{\rm op} &=& 1.8\times 10^{-37}\left( \frac{f}{1\,\mbox{Hz}}\right)^{2}.
\label{eq:Spmop}
\eea

In our prior studies \cite{Thorpe:2008wh, McWilliams:2009pe}, we neglected the $\Tbar$ channel due to inconsistencies that were observed between the analytic noise model for $S_{\Tbar}$ in (\ref{eq:Saet}) and the noise spectrum obtained from a time-domain simulation of $S_{\Tbar}$ performed with {\sc Synthetic LISA}. We have now verified that the two models are consistent for $f\ge 0.5\ \mbox{mHz}$. Where noted, we include contributions from the $\Tbar$ channel at these frequencies. We expect this to be a reasonably accurate approach since the response of the $\Tbar$ channel should vanish at low frequencies.

For the foreground of gravitational waves from 
unresolved compact binaries, we use the model 
developed in \cite{Timpano:2005gm}, by adding
\beq
S_{\rm gal} = [4\Phi\sin(\Phi)]^2 \,S_{\rm conf}
\label{eq:Sgal}
\eeq
to the expressions for $S_{\bar A, \bar E}$.
$S_{\rm conf}$ is a signal confusion noise estimate for a related measurement variable, taken from Eq.~14 of
\cite{Timpano:2005gm}. No gravitational-wave foreground is added to the $\Tbar$ channel since
$S_{\rm conf} \ll S_{\Tbar}$ at the frequencies where $\Tbar$ is included. 

\subsection{Error estimation}
\label{subsec:errorEst}

To approximate the achievable measurement accuracy for LISA, we apply
the Fisher matrix formalism. While not as robust or informative as more sophisticated methods such as Markov Chain Monte Carlo \cite{Cornish:2006ry}, the computational efficiency of the Fisher matrix approach is well suited for this type of performance estimation work. Potential pitfalls have been discussed in the literature \cite{Vallisneri:2007ev}.

The Fisher information matrix
is defined as
\beq
\Gamma_{ab} \equiv \bigg\langle \frac{\partial h}{\partial \Lambda^a}\, \bigg| \,
\frac{\partial h}{\partial \Lambda^b }\bigg\rangle,
\label{eq:fish}
\eeq
where $h$ is a channel of TDI strain, $a$ and $b$ are parameter indices, and $\langle\ldots |\ldots\rangle$ denotes a frequency-domain inner product weighted by the inverse noise spectral density $S_n(f)$ of the TDI output $n=\Abar,\,\Ebar,\,\Tbar$:
\beq
\langle x|y\rangle = 4\, \mathrm{Re} \int_{f_{\rm low}}^{f_{\rm high}} \frac{\tilde{x}^*(f)\tilde{y}(f)}{S_n(f)} \, .
\label{eq:innerprod}
\eeq
Here $f_{\rm low}$ is the low-frequency cutoff and $f_{\rm high}$ is
the half the sampling frequency described in 
Section~\ref{subsec:waveform}.
The individual Fisher matrices for each TDI channel to be included in the analysis are summed to form the total Fisher matrix.
 Throughout this work, we calculate the
parameter derivatives in \eqref{eq:fish} using second-order centered differencing (rather than one-sided as in \cite{McWilliams:2009pe}),
with a fractional step size $\varepsilon^a=\Delta\Lambda^a$.  We set $\Delta=10^{-4}$
for the coalescence time and $\Delta=10^{-6}$ for all other parameters.

The covariance matrix, $\Sigma^{ab}$,
is well approximated by the inverse of the Fisher matrix for signals with large SNR:
\beq
\Sigma^{ab} = \left(\Gamma^{ab}\right)^{-1} \left[1 + {\cal O}({\rm SNR}^{-1})\right],
\label{eq:cov}
\eeq
so that $\sigma^a \equiv \sqrt{\Sigma^{aa}}$ is the standard deviation of parameter $\Lambda^a$.
Because we use $\ln M$ and $\ln D_L$ as parameters,
the resulting uncertainties are fractional:
\bea
\sigma^{\ln M} &=& \sigma^M/M\,, \nonumber \\
\sigma^{\ln D_L} &=& \sigma^{D_L}/D_L\,.
\label{eq:lnsig}
\eea
The uncertainty of the other dimensionful parameter, $t_c$, is not expressed fractionally, as one is generally
interested in the absolute timing error, but we emphasize that one is free to make either choice.  We note that a poor choice of units for $t_c$ could cause the Fisher matrix to have a large dynamic range and be ill conditioned for inversion. We find that expressing $t_c$ in seconds yields computationally invertible Fisher matrices for the cases we consider.

\section{Sky localization}
\label{sec:res}

\begin{figure*}
\includegraphics[scale=.3]{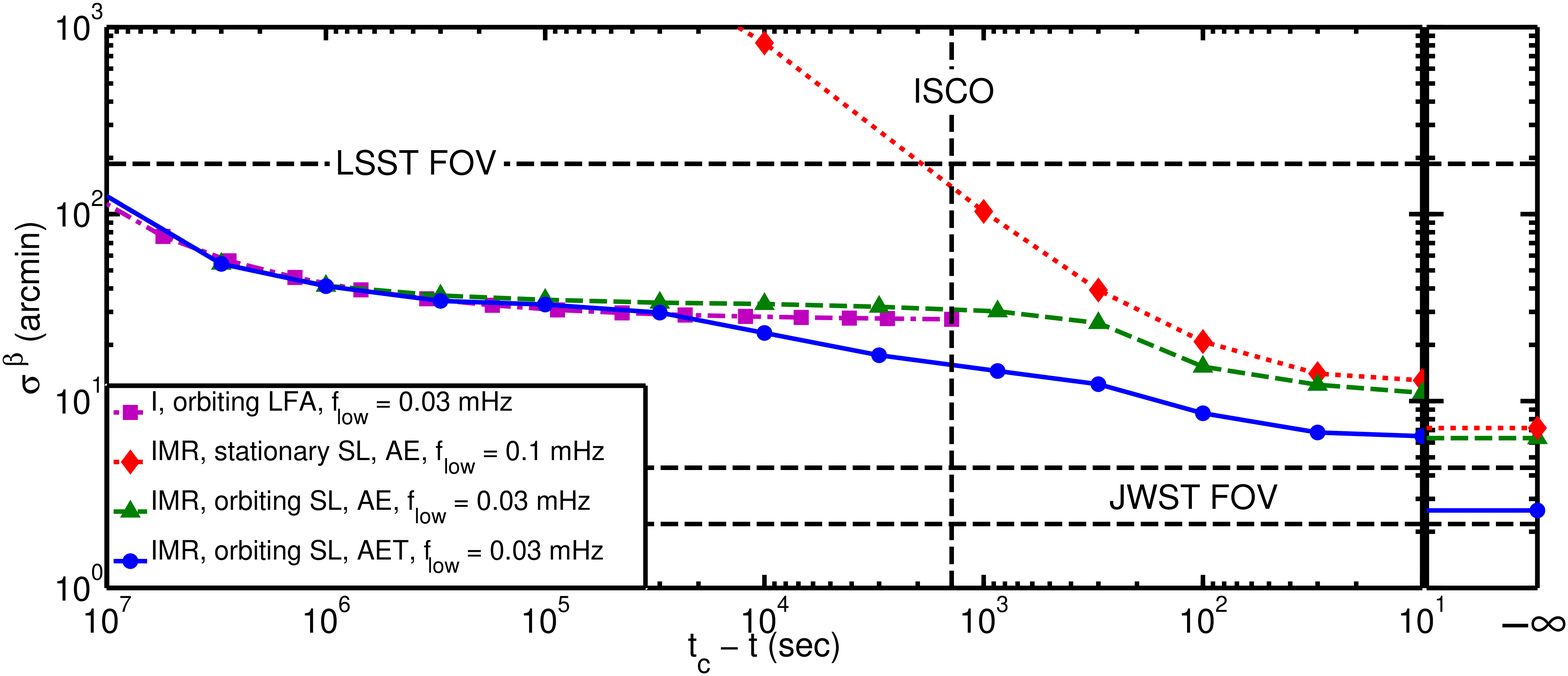}
\caption
{Comparison of latitude uncertainty estimates for an equal-mass, nonspinning MBHB with total mass $2\times 10^6\ M_{\odot}$ and redshift $z=1$.
The purple dot-dashed curve with square markers (based on \cite{Lang:2007ge, Lang:2008gh}) uses a post-Newtonian inspiral (I) waveform and a low-frequency approximation (LFA) to the detector response that includes LISA's orbit.  The red dotted curve with diamond markers (based on \cite{McWilliams:2009pe}) uses the complete inspiral-merger-ringdown (IMR) waveform and {\sc Synthetic LISA} (SL) for the response function but neglects LISA's orbit, all frequencies below $0.1\ \mbox{mHz}$, and the $\Tbar$ channel.  The green dashed curve with triangle markers uses the IMR waveform, the SL response, and orbital effects, but neglects the $\Tbar$ channel.  The blue solid curve with circle markers includes all effects: the IMR waveform, the SL response with orbits, and all three TDI channels.  The FOVs of LSST and JWST, as well as the location of the Schwarzschild ISCO, are shown for reference.}
\label{fig:sky}
\end{figure*}

As mentioned in Section~\ref{sec:intro}, the localization of a particular MBHB source on the sky is critical for enabling coordinated observations with electromagnetic observatories. One goal might be to locate the host galaxy in which the MBHB merger occurred. If the galaxy can be identified and the redshift obtained, it can be combined with the luminosity distance determined from the gravitational waveform for use as a cosmological probe \cite{Holz:2005df,Kocsis:2007yu}. 
The simplest way to identify a host galaxy would be for LISA to generate an error volume that only contained a single galaxy.  Unfortunately, even the most optimistic estimates of LISA's localization ability suggest that this will not generally be possible \cite{Kocsis:2007yu}. Identifying the host galaxy within LISA's error volume will then require other information, perhaps the detection of a transient electromagnetic signal associated with the merger itself. The nature of this possible signal is a subject of active research \cite{Schnittman:2010wy,Bloom:2009vx}, and it is unclear if it exists, when it might peak relative to the GW signal, and what its spectral content will be. 

The observational effort needed to search for an EM counterpart to an MBHB merger will depend on the ratio between LISA's sky localization error box and the field of view (FOV) of the EM asset used to perform the search. Roughly speaking, this ratio will determine the number of telescope pointings that are needed to cover the LISA error box.  While some counterpart scenarios involve afterglows that persist long after coalescence, others involve signals that peak at or before coalescence and may fade quickly.  For these signals, it is important to have some advance warning as to where on the sky (and when) the merger will occur, allowing for appropriate EM resources to be scheduled and pointed in advance.  The sky localization information produced by LISA's orbit accumulates with time over the months or years that the MBHB is observed, allowing for such advance notices to be issued.

Figure~\ref{fig:sky} summarizes various estimates for the development of
precision in ecliptic latitude over time for a LISA observation of
two equal-mass, nonspinning black holes with a combined rest mass $M_o=2\times 10^6\ \MSun$ at $z=1$. All of the curves show the median results from an ensemble of 100 systems with random sky locations and orientations.  Except where noted, the low-frequency limit of the inner product \eqref{eq:innerprod}, $f_{\rm low}$, was set to $0.03\ \mbox{mHz}$.  The precise placement of this cutoff and the model of the instrument noise between the cutoff and the lower limit of the ``official'' LISA measurement band at $0.1\  \mbox{mHz}$ can have a significant effect on the results \cite{Kocsis:2007hq,Kocsis:2007yu}.  The curves are all shown as functions of time before $t_c$, defined as the peak of the waveform.  In addition, we also show to the right the final error achieved with the complete MR signal.

The purple dot-dashed curve with square markers utilizes PN waveforms and a simplified model of the LISA instrument response known as the low-frequency approximation (LFA).  This approximation produces two channels of information which can be compared to the $\Abar$ and $\Ebar$ TDI channels discussed in Section~\ref{subsec:instrument}.  The estimate is generated by the same code used in \cite{Lang:2007ge,Lang:2008gh}, with the exception that here we ignore the effects of spin precession.  We include it primarily for comparison with our own method described in Section~\ref{sec:meth}.  Due to the breakdown of the post-Newtonian approximation, this curve is truncated at the time corresponding to the Schwarzschild innermost stable circular orbit (ISCO).  Therefore, this estimate is sensitive to sky localization information encoded in the waveform by LISA's orbital motion but is not sensitive to localization information resulting from inclusion of the MR waveform or from the high-frequency response of the instrument. 

The red dotted curve with diamond markers, based on our previous work \cite{McWilliams:2009pe}, uses the IRS-EOB waveform model, as described in Section~\ref{subsec:waveform}, and the {\sc Synthetic LISA} instrument response described in Section~\ref{subsec:instrument}.  Due to the aforementioned problem (now solved) of spurious information in the $\Tbar$ channel, only the $\Abar$ and $\Ebar$ TDI channels are included.  In addition, the instrument response uses a stationary, non-orbiting constellation lying in the ecliptic plane instead of the usual LISA orbits.  This estimate is therefore only sensitive to the sky localization information provided by the interaction of the MR signal with the high-frequency response of the detector.  Lacking the orbital contribution, we can safely remove a large portion of the inspiral by setting the low-frequency limit to $f_{\rm low} = 0.1\ \mbox{mHz}$ for this single estimate only.  This helps reduce computation time.

For times before the ISCO, the errors in the orbits-only estimate (purple, dot-dashed, squares) are far less than those in the stationary-LISA estimate (red, dotted, diamonds), suggesting that the orbital mechanism for sky localization dominates over these time periods.  The dotted curve only ``catches up'' to the dot-dashed curve $\sim100$ seconds before the merger.  However, by the time the full signal is considered, the stationary-LISA errors have become $\sim4$ times smaller than the orbits-only errors at ISCO. In other words, for nonspinning, equal-mass binaries in this mass range, observing the MR signal alone gives better localization information than observing the entire inspiral.

The remaining curves are new results, estimates which include both the complete IMR (IRS-EOB) waveform and a realistic, orbiting ({\sc Synthetic LISA}) model of the instrument. In addition, the low frequency cutoff has been restored to $0.03\ \mbox{mHz}$ in order to take advantage of the orbital modulation.  The green dashed curve with triangle markers includes only the $\Abar$ and $\Ebar$ TDI channels, while the blue solid curve with circle markers includes the $\Tbar$ channel as well.

The $\Abar\,,\Ebar$ estimate (green, dashed, triangles) tracks the orbits-only estimate (purple, dot-dashed, squares) at times well before merger, thus 
confirming that our calculation is consistent with that of Lang and Hughes \cite{Lang:2007ge,Lang:2008gh} in the inspiral regime.  It also tracks the stationary-LISA estimate (red, dotted, diamonds) as the MBHB approaches merger.  When the entire IMR waveform is included, the orbiting and non-orbiting $\Abar\,,\Ebar$ estimates differ by only $\sim10\%$.  The dashed curve contains both mechanisms for localizing a source (orbits and high-frequency detector response to merger-ringdown), but it ultimately does not do significantly better than the stationary-LISA response alone.  Our ability to localize sources well in advance of merger depends critically on the orbital information.  However, for the final achievable sky location accuracy in these systems, the inspiral is essentially irrelevant and only the MR signal need be taken into account.  

\begin{figure*}
\includegraphics*[scale=.16]{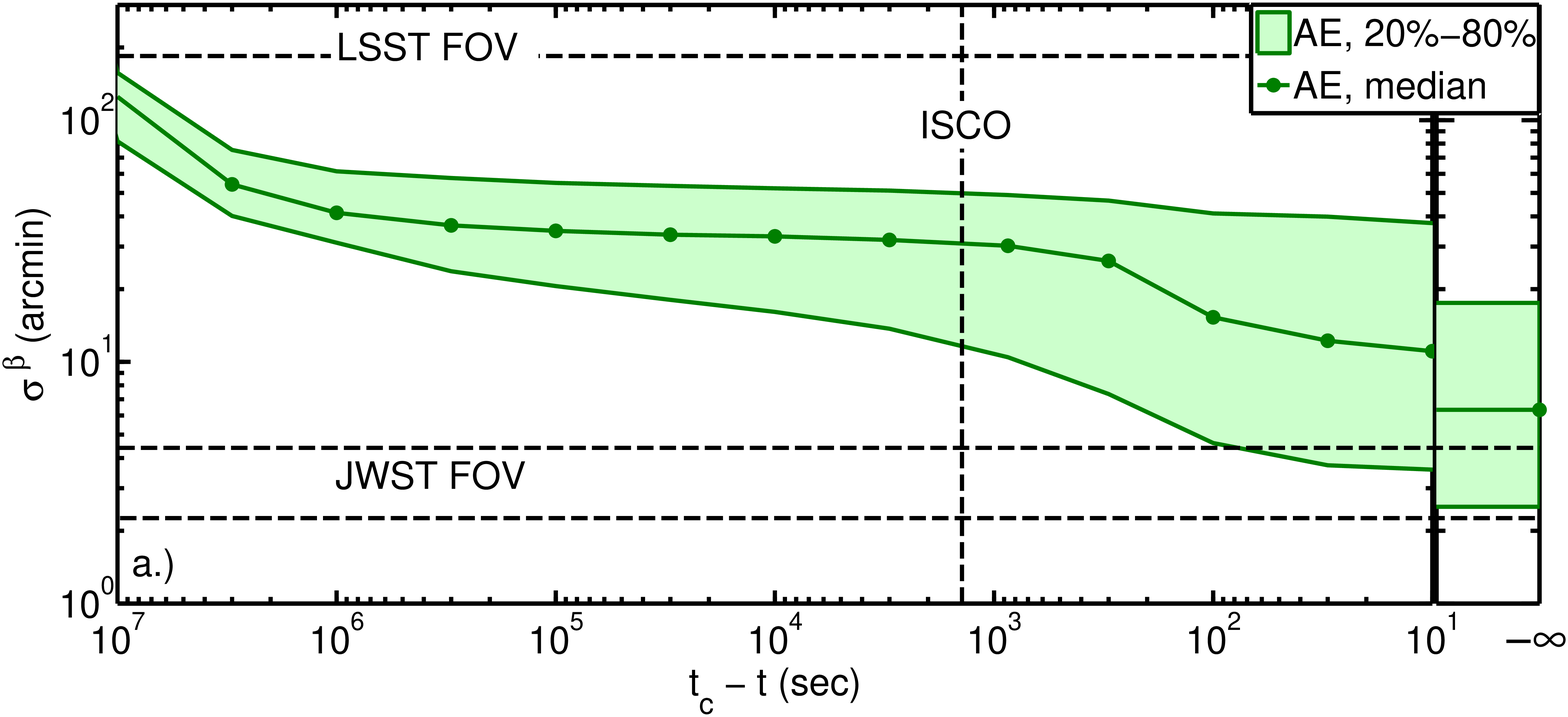}
\includegraphics*[scale=.16]{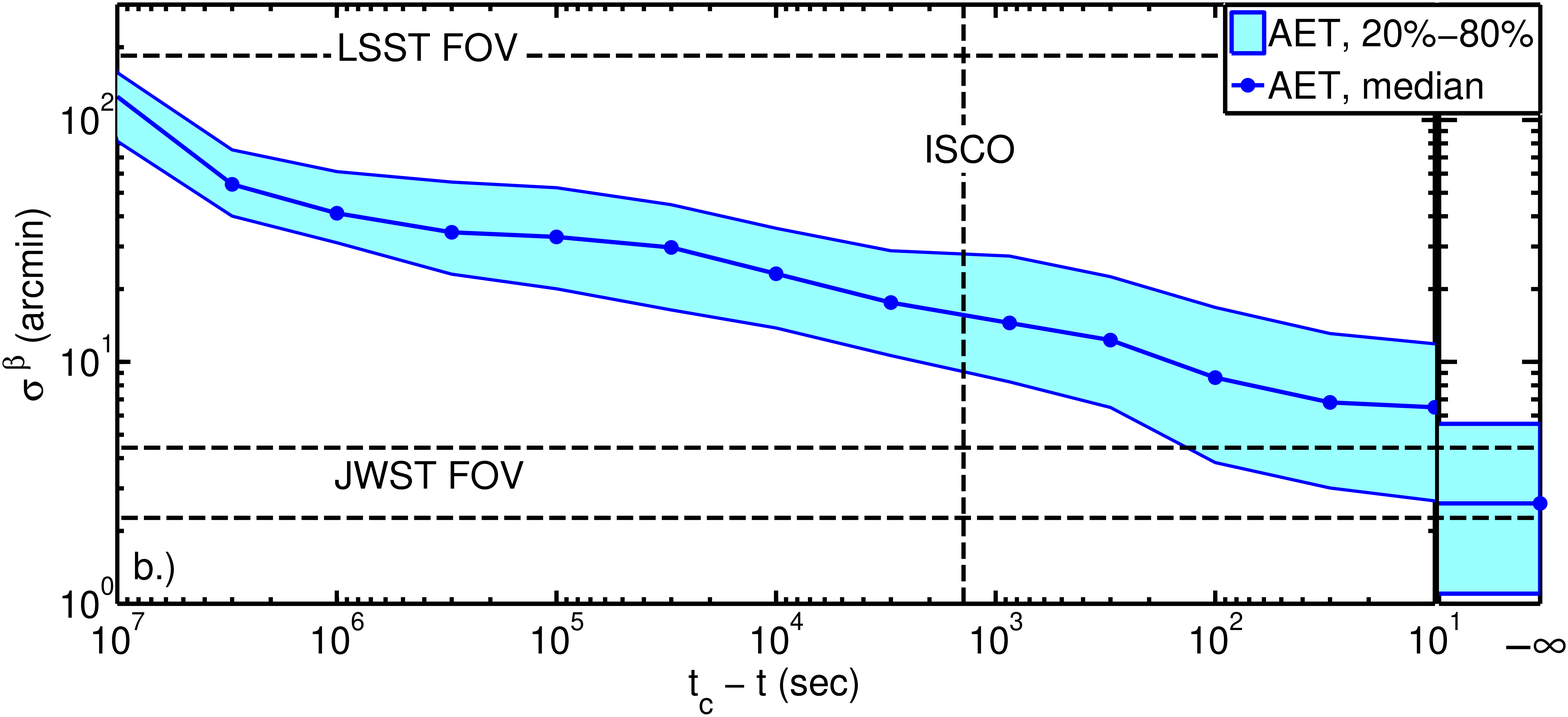}
\caption
{Latitude uncertainty, $\sigma^{\beta}$, for a $2\times 10^6\ M_{\odot}$ system at $z=1$.  The center line (with individual time samples marked) denotes the median error, while the shaded region depicts the middle three quintiles ($20^{th}-80^{th}$ percentile) of errors.  Both plots correspond to IMR signals with an orbiting LISA constellation.  However, panel (a) shows results for just the $\Abar$ and $\Ebar$ TDI channels, while panel (b) also includes the $\Tbar$ channel.}
\label{fig:lat}
\end{figure*}

\begin{figure*}
\includegraphics*[scale=.16]{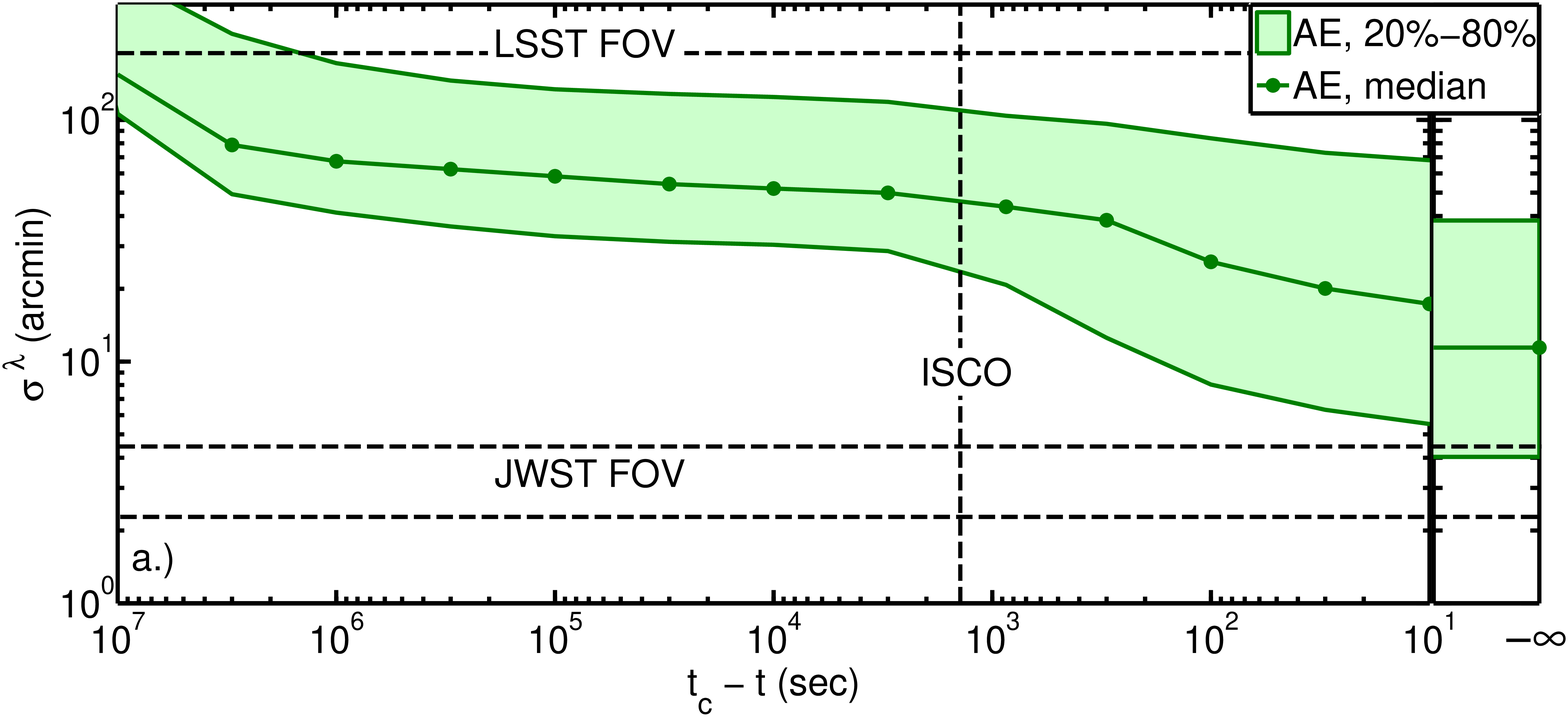}
\includegraphics*[scale=.16]{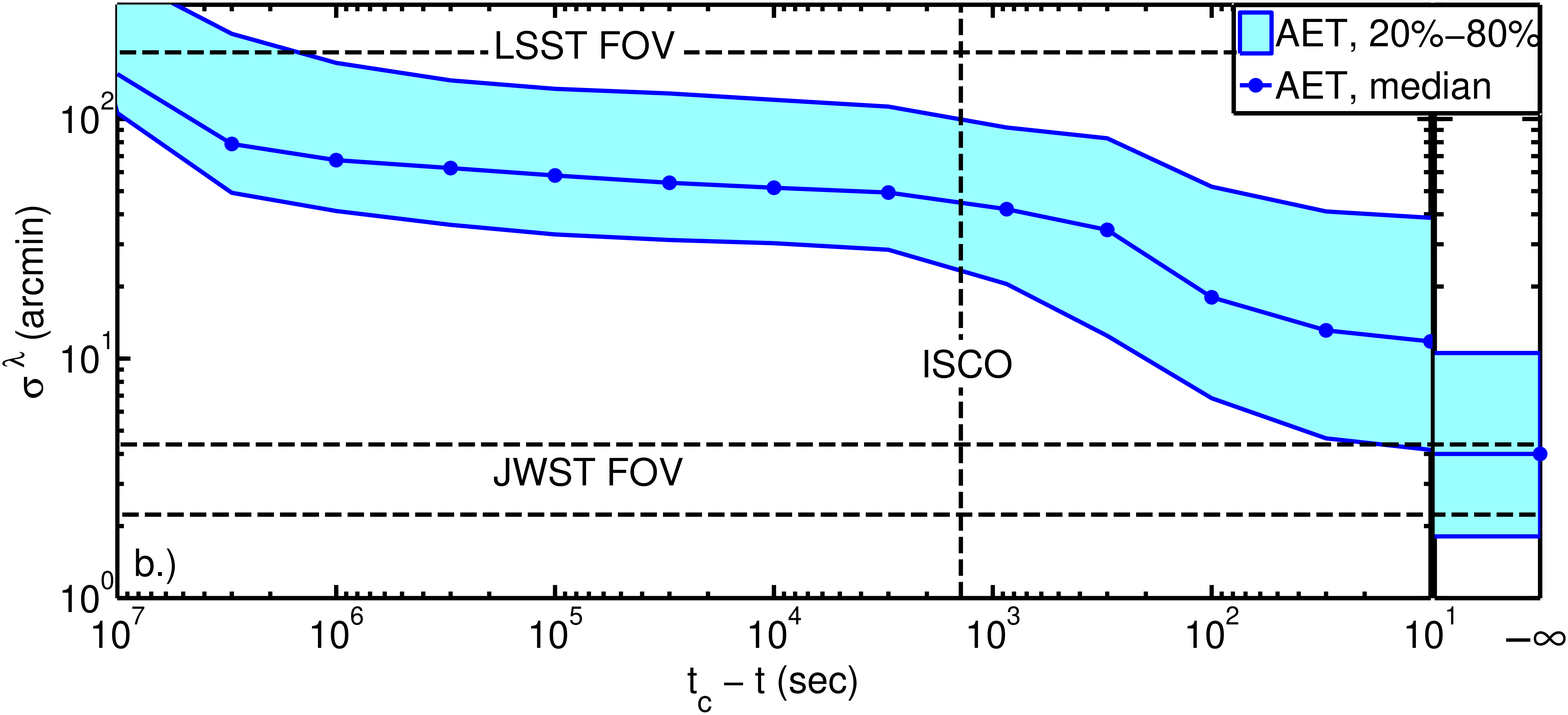}
\caption
{Same as Fig.\ \ref{fig:lat}, but for uncertainty in longitude, $\sigma^\lambda$.}
\label{fig:lon}
\end{figure*}

\begin{figure*}
\includegraphics*[scale=.16]{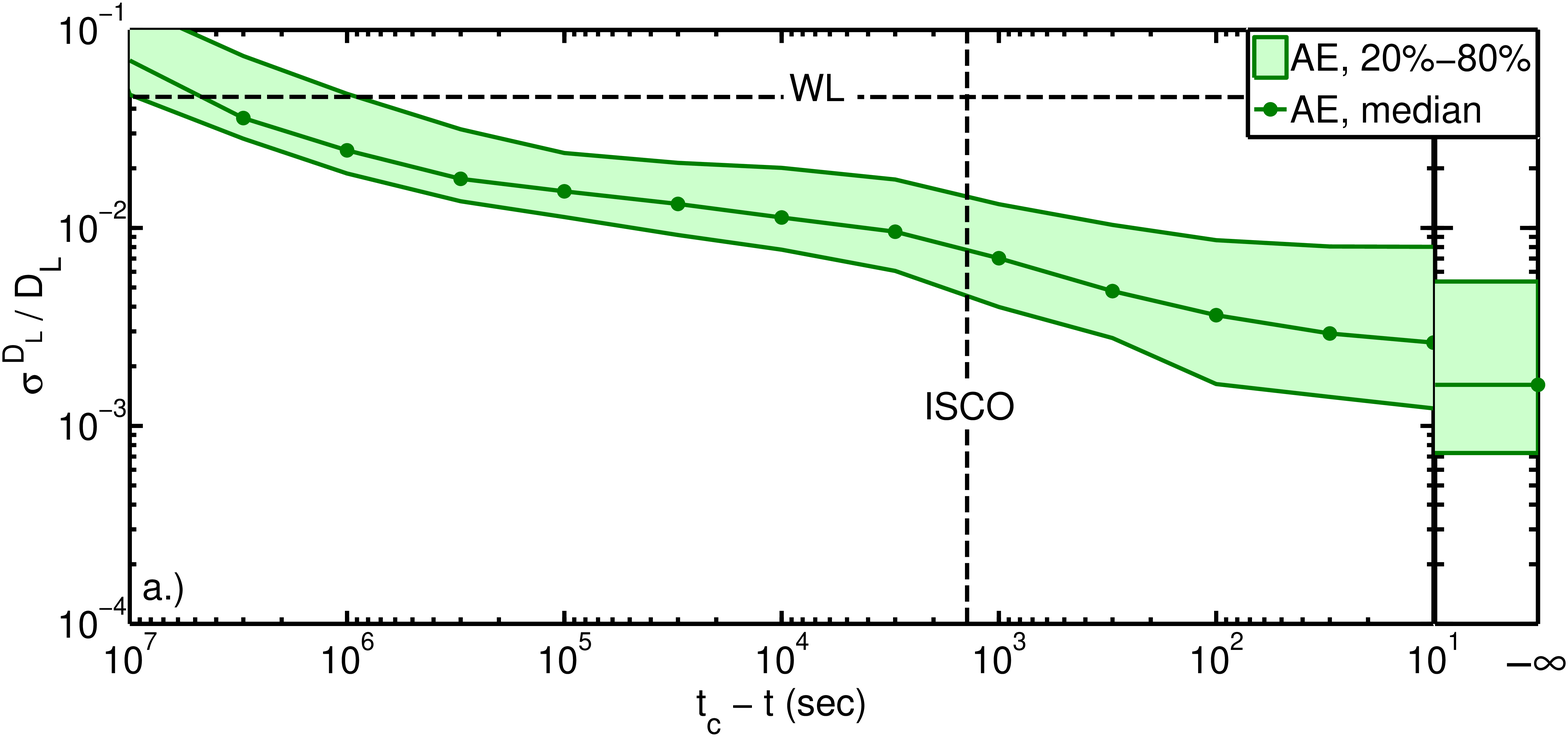}
\includegraphics*[scale=.16]{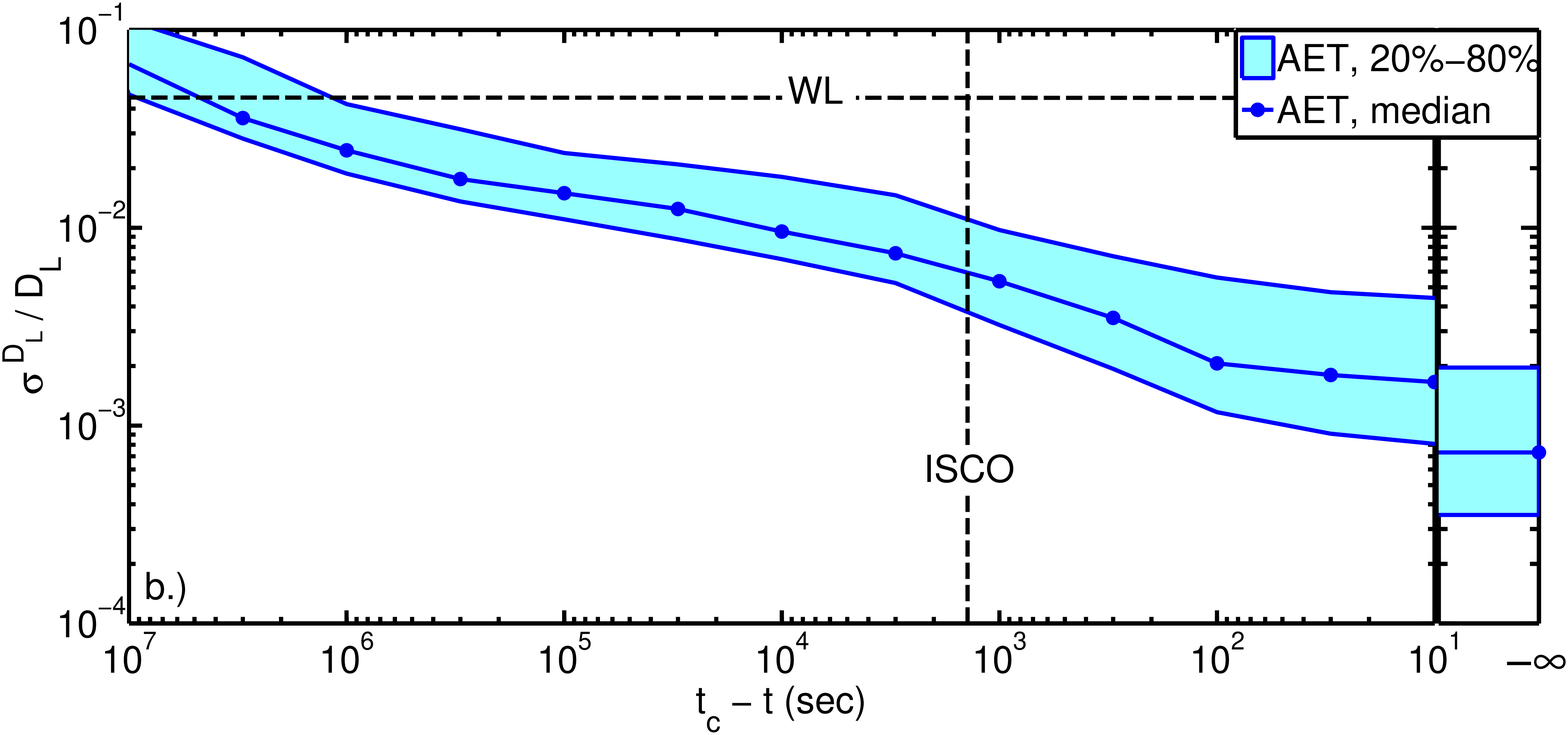}
\caption
{Same as Fig.\ \ref{fig:lat}, but for fractional luminosity distance uncertainty, $\sigma^{D_L}/D_L$.  
An estimate of the weak lensing limit (WL) \cite{HolzLinder} at $z=1$ is shown for reference.
Figs.~\ref{fig:lat}--\ref{fig:distance} collectively
show the evolution of the 3D error voxel.}
\label{fig:distance}
\end{figure*}

The estimate with all three TDI channels (blue, solid, circles) begins to diverge from the orbits-only estimate (purple, dot-dashed, squares) at earlier times than the $\Abar\,,\Ebar$ estimate and reaches a value at coalescence that is about half that of the $\Abar\,,\Ebar$ estimate. This represents the additional contribution of the $\Tbar$ channel, which is concentrated at high frequencies.  It is neglected implicitly in the orbits-only estimate through the use of the LFA and explicitly in the other two estimates.  While all three TDI channels are available in a fully operational, 6-link LISA, only two channels are available if one link is down \cite{Vallisneri:2008}.  Using the green dashed line as a proxy for a 5-link LISA and comparing to the blue solid line, we can see roughly how much the sky position determination is affected by a non-optimal detector.

Figure~\ref{fig:sky} also includes the FOV of two representative EM instruments that could be used for locating counterparts. The Large Synoptic Survey Telescope (LSST) \cite{LSST} has a FOV of $9.6\ \mathrm{deg}^2$, which we have modeled as a square aperture $\sim186\ \mathrm{arcmin}$ on a side, and is designed to rapidly and repeatedly scan large areas of the sky. The figure demonstrates that LISA will localize a typical equal-mass, nonspinning MBHB source of this mass and redshift within a single LSST field several months in advance of merger. This will allow multiple exposures of the target area to be made throughout the event in an effort to detect an optical transient. 

If a transient signal is not bright enough to be visible with a survey telescope such as LSST, then a more sensitive instrument with a narrower field of view may be required. In Fig.~\ref{fig:sky} we plot the $2.2\ \mathrm{arcmin}\times4.4\ \mathrm{arcmin}$ FOV of the NIRCam instrument on the James Webb Space Telescope (JWST) \cite{JWST} as an example of such an instrument. Our most complete estimate of LISA's localization capabilities 
(IMR waveform, orbiting detector with {\sc Synthetic LISA} response, all TDI channels) has a median latitude error of 
$2.6$ arcmin. This suggests that a single JWST exposure might be sufficient to cover the LISA error box for this system. In reality, a 
few exposures may be required to account for the shape of the LISA error box, its orientation with respect to the JWST FOV, and any operational constraints of JWST. The results in Fig.~\ref{fig:sky} suggest that unlike the situation with LSST, in which a number of pre-merger images could be obtained, a coordinated observation with JWST (or any other instrument with a similar FOV) will have to be a follow-up, as LISA requires the complete IMR waveform to provide precise localization information. 
For example, at ISCO, the best estimate for median latitude error exceeds the JWST FOV by roughly half an order of magnitude. 

All of the curves in Fig.~\ref{fig:sky} represent the median LISA latitude error for an ensemble of MBHB mergers.  Figures~\ref{fig:lat} and \ref{fig:lon} also show the distribution of both latitude and longitude errors, based on estimates with the complete IMR waveform and LISA's orbital modulation.  Separate panels show results including just $\Abar$ and $\Ebar$ and results including all three TDI channels.  In each panel, the solid line represents the median values (equivalent to what is plotted in Fig.~\ref{fig:sky}), while the shaded area shows the region spanned by the middle three quintiles of the distribution. The left panels ($\Abar$, $\Ebar$ only) show that the final LISA error box will lie within the FOV of JWST for systems in the $80^{th}$ percentile of sky localization precision.  That is, both latitude and longitude are measured to within $\sim4.4$ arcmin or better for the top $20\%$ of systems.   When $\Tbar$ is included, the situation is better still, with a full $50 \%$ of binaries in our ensemble fitting within the JWST FOV\footnote{These assertions are somewhat crude: The plot does not actually show whether {\it individual} binaries meet the criterion.  It is possible that some systems are well localized in latitude but poorly localized in longitude, and vice-versa.  A more careful check of the raw data verifies the numbers in the text.}.  These figures show just how important $\Tbar$ is for computing final localization errors.  Not only do the medians drop for both parameters, but the width of the distribution decreases as well.  Although a 5-link and 6-link LISA are indistinguishable at early times, complete functionality can have important effects for the full signal. 

For completeness, in Fig.~\ref{fig:distance} we show errors in luminosity distance, $\sigma^{D_L}/D_L$, for IMR signals with orbital modulation.  Distance is the third axis of the 3D localization ``voxel.''  Given an assumed cosmological model, a measurement of distance is equivalent to a measurement of redshift.  This approximate redshift can help decide between potential host galaxies located within the 2D error box.  We see that the median distance error drops below a percent within the last day, reaching a minimum of less than $0.1\%$.  Unfortunately, this ``intrinsic LISA error'' is not the limiting uncertainty for measuring luminosity distance.  Weak lensing from matter between the source and the detector will slightly magnify or demagnify the GW signal in an unknown way, adding significant error to the distance measurement.  The line marked ``WL'' on the figure is an estimate \cite{HolzLinder} of $\sigma^{D_L}/D_L\sim4.4\%$ from weak lensing at $z = 1$.  The intrinsic LISA error crosses the WL error over a month before merger; shortly after that time, the information on $D_L$ provided by LISA becomes irrelevant unless weak lensing error can be mitigated significantly \cite{WL}.
   
\section{Contribution of mergers to parameter estimation}
\label{sec:discussion}

The results presented in the preceding section clearly demonstrate that observing the MR portion of an MBHB signal can provide a significant amount of sky localization information, despite the fact that there is little or no modulation imposed by the LISA orbits. In this section, we explore the nature of this information.  Throughout this section, we always use the complete IMR waveform with orbital modulation and all three TDI channels.

\begin{figure}
\includegraphics[scale=.27]{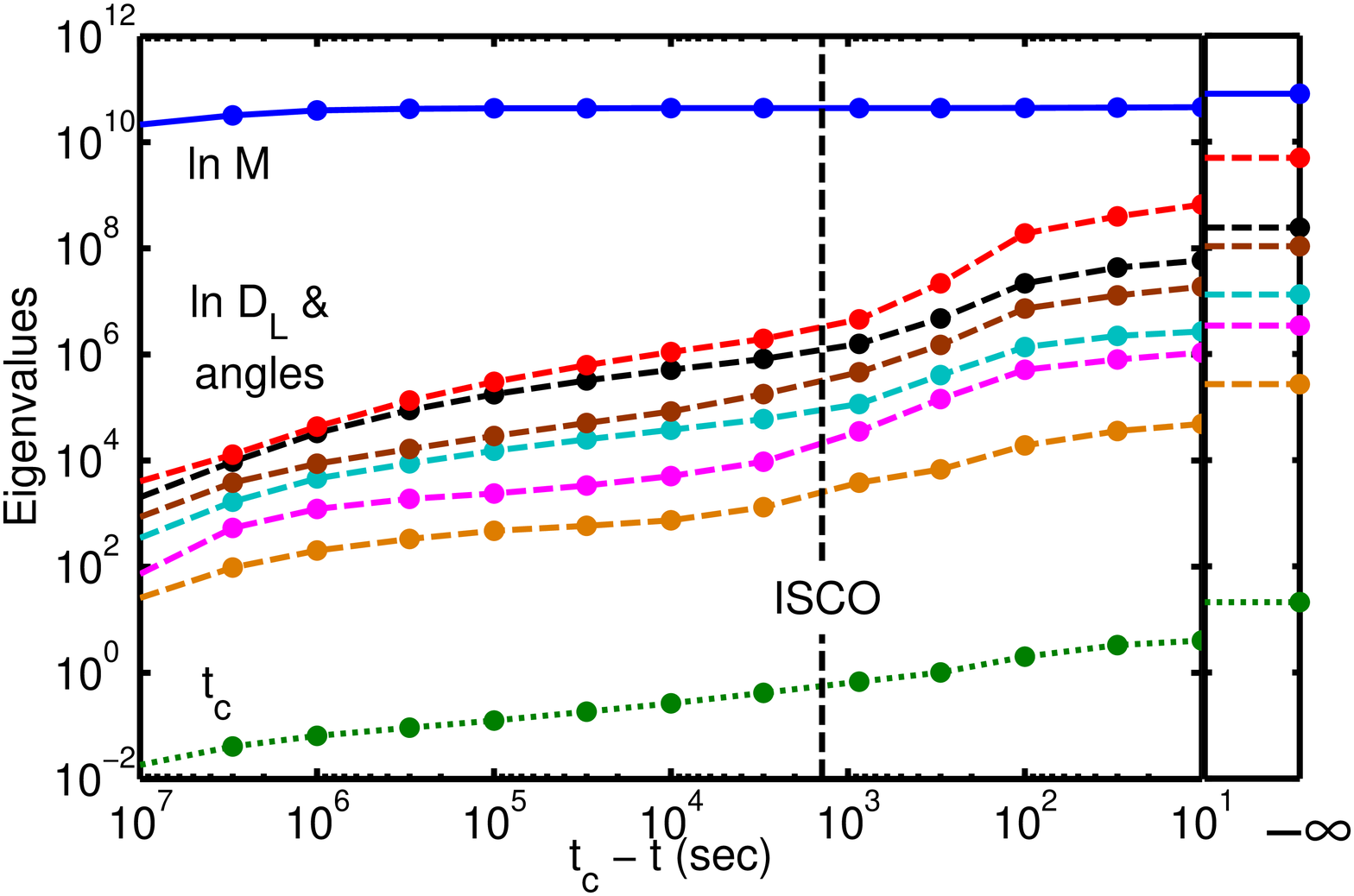}
\caption
{Evolution of the median Fisher matrix eigenvalues with time for the inspiral, merger, and ringdown of an ensemble of equal-mass, nonspinning MBHBs with $M_o = 2\times 10^6 \MSun$ at $z=1$ observed by an orbiting, six-link LISA via the $\Abar$, $\Ebar$, and $\Tbar$ TDI channels.}
\label{fig:eigenvalues}
\end{figure}

One question of interest is whether the information provided by the MR signal is tightly coupled to just one or two of the elements in $\Lambda^a$ or if it generally improves the knowledge of all parameters. Figure~\ref{fig:eigenvalues} shows the time evolution of the median eigenvalues of the Fisher matrix for the ensemble of systems studied in Section~\ref{sec:res}.  
At each time step, $\Gamma_{ab}$ was calculated using the $\Abar$, $\Ebar$, and $\Tbar$ observables for every case in the ensemble using the procedure described in Section~\ref{sec:meth}. These matrices were used to compute the eigenvalues for every case in the ensemble, and the median values are plotted. A similar procedure was used to compute the median eigenvectors corresponding to these eigenvalues.

As (\ref{eq:cov}) indicates, the Fisher matrix is a measure of inverse error or ``information.''  The increasing trends in the curves in Fig.\ \ref{fig:eigenvalues} represent information accumulating in the LISA observables as the signal evolves, just as the decreasing trends in Figs.\ \ref{fig:sky}-\ref{fig:distance} represent a decrease in measurement error. 
The largest eigenvalue in Fig.\ \ref{fig:eigenvalues} (solid curve) corresponds to an eigenvector that mostly consists of one parameter, $\ln M$. The majority of the information in this eigenvalue accumulates early in the inspiral phase of the signal. From ISCO through the end of the signal, the eigenvalue roughly doubles, with most of the increase occurring very near the time of coalescence. This is consistent with previous results that show that $\ln M$ is primarily determined during the inspiral phase \cite{Cutler:1994ys} and that the MR signal provides little additional information \cite{McWilliams:2009pe}. By comparison, the increase in ${\rm SNR}^2$, which one might naively expect to be proportional to Fisher matrix eigenvalues, is $\sim100$ over this period.  

The band of eigenvalues in the middle of Fig.~\ref{fig:eigenvalues} (dashed curves) represent a mix of the five angular parameters in $\Lambda^a$ and $\ln D_L$. The increase in these eigenvalues from ISCO through ringdown is much more significant than that for $\ln M$, ranging from two to three orders of magnitude. 

The smallest eigenvalue (dotted curve) corresponds to an eigenvector that mostly consists of $t_c$. The small size of this eigenvalue is partly due to the intrinsic scale of the parameter, which was the only one in $\Lambda^a$ to be expressed in physical units (seconds). It increases by a factor of $\sim30$ between ISCO and the end of ringdown.

The fact that all of the eigenvalues increase post-ISCO indicates that the MR portion of the waveform, coupled with the high-frequency response of the LISA instrument, provides information in all directions of parameter space.  We might have expected that the merger, with its pronounced peak, 
would provide a very accurate measurement of $t_c$ and only affect other parameters
by decreasing their degeneracy with $t_c$.  However, Figure~\ref{fig:eigenvalues} demonstrates that this is not the case.  If we were to assume that sky localization results primarily from the modulation imposed by LISA's orbit, 
then we would expect that the merger cannot provide any intrinsic information about sky position, since the
LISA constellation remains effectively stationary during merger. Figure~\ref{fig:eigenvalues} demonstrates that the interaction between
the MR signal and the complex response of the LISA instrument provides information about every parameter independently.  This is in addition to any contribution from degeneracy breaking which may occur.

\section{Summary and Conclusions}
\label{sec:conc}

We have presented the first estimates of LISA's ability to measure the source parameters of nonspinning, equal-mass MBHBs that include a complete inspiral-merger-ringdown waveform, realistic LISA orbits, and the full high-frequency response of the LISA instrument. For systems with $M_o = 2\times 10^6\ \MSun$ at $z=1$, we find that LISA can locate the source on the sky with an error of $\sim3$ arcminutes in the median case at the end of the MBHB event. We have also studied how the sky localization information for these systems evolves with time and find that, at ISCO, the sky localization errors are typically tens of arcminutes.

Inspired by the potential for multi-messenger astronomy enabled by combined observations of MBHB mergers with gravitational-wave detectors and electromagnetic telescopes, we have compared the source-localization capabilities of LISA with the FOVs of two future telescopes: LSST and JWST. We find that most systems in our ensemble are located within the LSST FOV several months prior to coalescence.  Localization within the smaller FOV of JWST requires information from the MR waveform.

While the results in this paper focus on a particular instrument (LISA) and a particular source class (equal-mass, nonspinning, $M_o = 2\times10^6\ \MSun$, $z=1$) they can provide insight for other scenarios as well.  For example, proposals for lower-cost variants of LISA often involve a constellation with shorter arm lengths \cite{Stebbins:2009kz}. This has the general effect of shifting the response of the instrument to higher frequencies.  We have asserted that the improvement in parameter estimation that we observe post-ISCO is due in part to the complexity of the high-frequency response of the instrument.  If this is true, shortening the arms could have a significant impact on the ability to localize systems in the mass range studied. Specifically, the improvement in sky localization from observing the merger-ringdown of a system with $M_o\sim 10^6\ \MSun$ may not be as dramatic as that presented in Section~\ref{sec:res}. At the same time, the decrease in strain sensitivity associated with shorter arms would reduce the duration of the inspiral signal that is observed, further degrading measurement performance for these systems. For systems with larger $M_o$, the situation would be worse. 

For lower-mass systems, shortening the arms (if the interferometric measurement system is not simultaneously degraded) may actually increase the benefits of observing the merger by increasing the SNR for the merger-ringdown portion of the signal. Since these lighter systems have higher merger frequencies, they will still experience a complex response in the detector, even with shorter arms. Determining whether this effect can overcome the lower intrinsic SNR of less massive systems will require a more detailed calculation. 

Finally, we note that while the addition of the merger-ringdown signal and a complete detector response is an important step towards realizing the full potential of LISA and other space-based gravitational-wave detectors, we still do not have a complete picture.  Other potential sources of information in the MBHB waveform, particularly spin and the associated precession effects, have been demonstrated to significantly improve LISA's localization ability in the inspiral-only case \cite{Lang:2006bz}. It remains to be seen what impact combining spin precession and merger will have on LISA sky localization estimates.

\vspace{4.5mm}

\hspace{0.85in}
{\bf Acknowledgments}

\vspace{3.5mm}

We thank Tuck Stebbins for helpful discussions and Bernard Kelly for his thorough review of the manuscript.  We acknowledge the support of NASA grant 08-ATFP08-0126.
RNL was supported by an
appointment to the NASA Postdoctoral Program at the Goddard Space
Flight Center, administered by Oak Ridge Associated Universities through
a contract with NASA.  The simulations were carried out using resources from
the NASA Center for Computational Sciences (Goddard Space Flight Center).

Copyright (c) 2011 United States Government as represented by the Administrator of the National Aeronautics and Space Administration. No copyright is claimed in the United States under Title 17, U.S. Code. All other rights reserved.

\end{document}